\documentclass[11pt]{article}
\usepackage{graphicx} 
\usepackage{caption}
\usepackage{amsmath}
\usepackage[T1]{fontenc}
\usepackage[utf8]{inputenc}
\usepackage{authblk}
\usepackage{hyperref}
\usepackage{multirow}

\usepackage{mathtools}
\usepackage{amsfonts}

\DeclarePairedDelimiter\floor{\lfloor}{\rfloor}
\newtheorem{theorem}{Theorem}
\newtheorem{corollary}{Corollary}

\title{Discovering Multiple Phases of Dynamics by Dissecting Multivariate Time Series}
\author[1]{Xiaodong Wang}
\author[1*]{Fushing Hsieh}
\affil[1]{Department of Statistics, University of California, Davis.}
\date{} 

\begin{document}

\maketitle

\section*{Abstract}


We proposed a data-driven approach to dissect multivariate time series in order to discover multiple phases underlying dynamics of complex systems. This computing approach is developed as a multiple-dimension version of Hierarchical Factor Segmentation(HFS) technique. This expanded approach proposes a systematic protocol of choosing various extreme events in multi-dimensional space. Upon each chosen event, an empirical distribution of event-recurrence, or waiting time between the excursions, is fitted by a geometric distribution with time-varying parameters. Iterative fittings are performed across all chosen events. We then collect and summarize the local recurrent patterns into a global dynamic mechanism. Clustering is applied for partitioning the whole time period into alternating segments, in which variables are identically distributed. Feature weighting techniques are also considered to compensate for some drawbacks of clustering. Our simulation results show that this expanded approach can even detect systematic differences when the joint distribution varies. In real data experiments, we analyze the relationship from returns, trading volume, and transaction number of a single, as well as of multiple stocks in S$\&$P500. We can successfully not only map out volatile periods but also provide potential associative links between stocks.


\section{Introduction}

It has received increasing research interests and attention in studying nonlinear stochastic dynamics in quantitative finance. Researchers and practitioners have realized that it is rather important to understand the joint behaviors of multiple aspects of one single stock or asset as well as one common aspect of multiple assets. One dynamic issue that has been making quantitative finance experts wondering even up to now is the joint dependence among returns, trading volume, and transaction numbers \cite{Ying, 24}. Another well-known dynamic issue is about how volatility clustering comes to exist and where to look for it \cite{Ding}. Since volatility is measured by conditional variance and it changes over time for one single stock or asset, how to compute and visualize volatility in concert with a form of clustering to a great extend is still mysterious. Computational and data-driven approaches for both issues are not yet well established or reported in the literature.

For instance, GARCH models have been proposed to study and to model the time-varying volatility of asset returns \cite{1}, and their variants have been extended to multivariate time series cases by modeling the correlation dynamics \cite{2,3,4}. However, they require too many parameters and a large collection of prior knowledge about the dynamic structure. Such modeling and required structures make the model interpretation rather complicated. A more effective methodology was proposed to incorporate realized volatility \cite{5,6} and realized covariance \cite{7,8}. However, its results could be biased due to noises’ hard to be accommodated nature. Further, often a long time window is usually required implement such a methodology.

Recently, a data-driven approach named Hierarchical Factor Segmentation(HFS) is developed by characterizing volatility fluctuation directly \cite{9}. HFS, to some extent, is similar to the regime-switching model advocated by \cite{10,11}. Both likewise assume regime-switches being somehow away from the beginning and ending time points of the involved time span. HFS computationally attempts to detect all time-points, at which the dynamics phase shifts from one episode to another by revealing distinct dynamic behavior. Its chief computing device tracks the recurrence of ``extreme’’ events, i.e. large returns, defined by a chosen threshold. Consequently, latent regions with different event-intensities are segmented. Via this way, dynamic tail behaviors are successfully discerned. Along the direction of threshold choice, HFS was extended to study the empirical tail distribution by applying a series of thresholds in \cite{12}. 

Compared with region-switching models, HFS takes advantage of offline analysis to decode dynamics patterns without assuming any underlying distribution or Markovian structure. So HFS is in the category of nonparametric change point detection in time series. Nonparametric change point detection has a wider range of applications than parametric \cite{13}. Characteristically, it often relies heavily on the estimation of density functions \cite{14}, see details in a recent survey being available in \cite{15}. The key difference between HFS and the change point approach, in general, is that we assume the underlying distributional changes at a certain point and may come back in afterward in a recurrent fashion, which makes more sense in the case that volatility clustering comes and leaves recurrently in financial data. 

So far, the nonparametric approach in discovering the recurrent switch patterns underlying multivariate time series is still scarce. One underlying reason is attributed to the fact that the nonlinear dependence among them is the key and necessary knowledge. Thus, missing or lacking such knowledge underlying all involving time series or processes has become a barrier that hinders the potential research advances in this direction. Such dependence needs to be measured based on the latent phases revealed from each single process separately \cite{12,16}. However, beyond the multiplicity and complexity of global dynamic patterns, the nonlinear dependence can be easily overwhelmed by the integrated microstructure noises.  In this paper, we extend the idea of HFS to discern the temporal switching patterns underlying a collection of assets. This extended computational approach proceeds in three steps. Firstly, a chosen $\mathbb{R}^p$ dimensional region is created based on observed time series data and then partitioned into $B$ subareas. Upon each subarea, its chronological emergence is tracked along the temporal axis of the involved time series. Secondly, the limiting distribution of recurrent time between successive events according to each subarea-specific chronological emergence is analyzed. Then, a confusion matrix is constructed by stacking $B$ estimated permission rate vector resulted from each subarea. Lastly, clustering analysis is applied to group similar time points as if they are sharing the same phases of the dynamics. Via such clustering, the dynamic patterns of hidden phases are revealed by the cluster index. This is the fundamental idea underlying our proposed methodological extension of HFS.

The paper is organized as follows. In Section2, we introduce the asymptotic theory for a homogeneous recurrent time distribution and then describe the HFS algorithm, which can be applied when distributions switch temporally. In Section3, we proposed our main method in segmenting multivariate processes. In Section4, feature-weighting techniques of clustering are proposed for choosing potentially informative ``extreme’’ events. Simulation experiments and real data analysis on multiple time series of one stock and multiple stocks in S$\&$P500 index are performed in Section5 and Section6, respectively.

\section{Recurrent Time Distribution}

\subsection{Homogeneous Time Series}

Given a large data of stock price at an even interval of time and its consequential calculated returns with length $N$, we can encode the continuous time series into a 0-1 binary sequence of length $N$ where 1 indicates an observation of a rare event, and 0 otherwise. In such one-dimensional stock time series, the rare event is defined by extremely large values of absolute stock returns, so that the binary sequence can represent the frequency of return volatility. Note that a period with a high-frequent appearance of 1's may indicate a volatility clustering. 

As advocated by the Black-Scholes model, stock's return stochastic process is often modeled by geometric Brownian motion, or more generally geometric Levy processes. The returns are i.i.d. or exchangeable under the model assumptions. It motivates an invariance theorem for the waiting time between successive extreme large returns. If we look at the time of observing a successive 1's under the assumption of exchangeable returns, it was proved that the waiting time is asymptotically independent and its finite empirical distribution converges almost surely to a geometric distribution \cite{17}.

Consider a $N$-length series of stock returns $\{X_t\}_{t=1}^N$. Suppose $M$ out of $N$ objects are selected randomly as 1's, and the unseleted $N-M$ as 0's. Denote the recurrent time of two successive 1's as $R^N$, so there obtained $M+1$ recurrent time sequence $R^N_1, R^N_2, ..., R^N_{M+1}$. Assume the waiting time can be 0 if two 1's appear consecutively, $R^N_1=0$ if $X_1=1$, and $R^N_{M+1}=0$ if $X_N=1$. Due to the exchangeability assumption of $\{X_t\}_{t=1}^N$, $R^N_1, R^N_2, ..., R^N_{M+1}$ are exchangeable as well. 

\begin{theorem}\label{th1}
If $N \rightarrow \infty$ and $M \rightarrow \infty$ in a way such that $\frac{M}{N}\rightarrow p \in (0,1)$, then, for any $t \ge 1$,
\begin{equation}
(R^N_1, R^N_2, ..., R^N_t) \xrightarrow{d} (R_1, R_2, ..., R_t)
\end{equation}
where $(R_1, R_2, ..., R_t)$ are independent and identically geometric distribution with parameter $p$.
\end{theorem}

The proof sees Theorem 2.1 in \cite{17}. When $N$ and $M$ go to infinity in a way that $M \sim Np$, the recurrent time becomes asymptotically independent and converge to a geometric distribution with parameter $p=M/N$. 

In one-dimensional stock returns, a fixed proportion from all the time points can be selected as events of interest. For example, $\alpha$ and $\beta$ quantile is set to cut the lower and upper tail of the distribution, where $0<\alpha < 0.5 < \beta <1$ and $\alpha + (1 - \beta) < 1$. So that all time stamp has the same probability $p= \alpha + (1-\beta)$ to be marked as 1. A excursion process is defined by,
\begin{equation}\label{eq:cut}
E_t = \begin{cases} 1 & \quad X_t\le \alpha\textit{-quantile}, \,\, X_t\ge \beta\textit{-quantile}\\
0 & \quad Otherwise \end{cases}
\end{equation}
where $\{E_t\}_t$ is the resultant 0-1 sequence after labeling absolute large return as 1 and 0 otherwise.

However, the exchangeability of successive returns is easily violated due to the well-known stochastic volatility in finance. Returns should only be considered exchangeable locally, and rapid time-varying volatility is evidently observed \cite{18}.

\subsection{Region Switching Model}

Complicated models are investigated to evaluate the mechanism of stock return. Markov property, more or less, plays a significant role in modeling time-varying volatility. Instead, Hierarchical Factor Segmentation(HFS) is proposed to search for alternating hidden regions without assuming any Markov property \cite{9}. The idea is to admit distributional heterogeneity embedded behind the distribution of stock returns. After encoding the time series in the same way described above, HFS is implemented to label each time point by an index of hidden regions. By assuming that the conditional distributions are exchangeable within a hidden region, a corollary for heterogeneous time series is shown as a direct consequence of Theorem \ref{th1}.

Assume there are only $k$ hidden regions, denoted by $S_1, S_2, ..., S_k$. Select a fix size of samples $M$ from all the samples. Denote the sample selected from region $S_j$ having size $M_j$, for $j=1,2,...,k$. So, $\sum_{j=1}^{k} M_j =M$.

\begin{corollary}
If $N \rightarrow \infty$ and $M \rightarrow \infty$ in a way such that $\frac{M_j}{N} \rightarrow p_j \in (0,1)$, for $j=1,2,...,k$, then, for any $t \ge 1$,
\begin{equation}
(R^{N}_1, R^{N}_2, ..., R^{N}_{t} | S_j) \xrightarrow{d} (R_1, R_2, ..., R_{t} | S_j)
\end{equation}
where $(R_1, R_2, ..., R_{t} | S_j)$ are independent and identically geometric distribution with parameter $p_j$ given hidden region $S_j$.
\end{corollary}

Moreover, by further assuming identicality of conditional distribution given a hidden region, i.e. the cumulative distribution function denoted by $F_j$ given region $S_j$, with an appropriate choice of $\alpha$ and $\beta$ advocated in \eqref{eq:cut}, ratio $\frac{M_j}{N}$ converges to a constant almost surely,

\begin{equation}
\frac{M_j}{N}\rightarrow p_j= (\int_{-\infty}^{\alpha\textit{-quantile}} + \int_{\beta\textit{-quantile}}^{\infty}) dF_j
\end{equation}

HFS generates all possible decoding candidates with only a few tuning parameters. For example, two parameters are enough for generating a sequence with 2 alternating hidden regions. $k+1$ parameters are needed to generate $k$ hidden regions. Thus, an exhausted searching algorithm can be implemented to find the global optima. To measure goodness-of-fit for a potential hidden state sequence, model selection is done by fitting geometric distribution within each hidden region, while penalizing the total number of switching regions as the model complexity. Information criteria AIC or BIC can be utilized for this purpose. Parameter $p_j$ is estimated by MLE $\hat{p_j}=M_j/N$. So, the resultant loss function can be written as,

\begin{equation}
Loss(\theta) = -2 \sum_{j=1}^{k} [\sum_{t \in S_j^{\theta}}{E_t}log\hat{p_j} + \sum_{t \in S_j^{\theta}}{(1-E_t)}log(1-\hat{p_j})] + \phi(N)Q_k
\end{equation}
where $E_t$ is a 0-1 discrete process after applying binning strategy \eqref{eq:cut}; $k$ is the number of hidden states; $Q_k$ is the total number parameters from $k$ conditional geometric distributions; $\phi(N)=2$ for AIC or $\phi(N)=log(N)$ for BIC; $\theta$ is the parameter involved in HFS.

For the completeness of this section, HFS in partitioning a sequence with two hidden states is shown in Algorithm~1. Details about decoding multiple hidden states, and how to choose the number of hidden states $k$ are investigated in \cite{12}. Denote the recurrence time between successive 1's as $\{R_t\}_{t}$, and parameter vector as $\theta=(T,T^{*})$.

\noindent\rule{12.5cm}{0.8pt}\\
\textbf{Algorithm~1} Hierarchical Factor Segmentation(HFS)\\
\rule{12.5cm}{0.4pt}\\
Initial: an empty time set $S_0$\\
Input: a sequence of recurrence time $\{R_t\}_{t}$\\
1. Transform $\{R_t\}_{t}$ sequence into a 0-1 digital strings $\{E^{*}_t\}_{t}$ via a second-level coding scheme:
\[ E^{*}_t = \begin{cases} 1 & \quad R_t\ge T\\
0 & \quad Otherwise \end{cases}\]
\noindent2. Upon code sequence $\{E^{*}_t\}_{t}$, take code digit 1 as another new event and recalculate the event recurrence time sequence $\{R^{*}_t\}_{t}$.\\
3. Loop: cycle through every $R^{*}_t$\\
\indent if $R^{*}_t \ge T^{*}$, then add its corresponding time point $S^{'}$ into set $S_0$
$$S_0=S_0 \bigcup S^{'}$$
\indent where set $S_0 \subset \{1,...,n\}$ and could be empty.\\
4. Define the other internal state $S_1=\{1,...,n\}\textbackslash S_0$.\\
\rule{12.5cm}{0.8pt}\\

\begin{figure}[!h]
\centering
\includegraphics[width=5.2in]{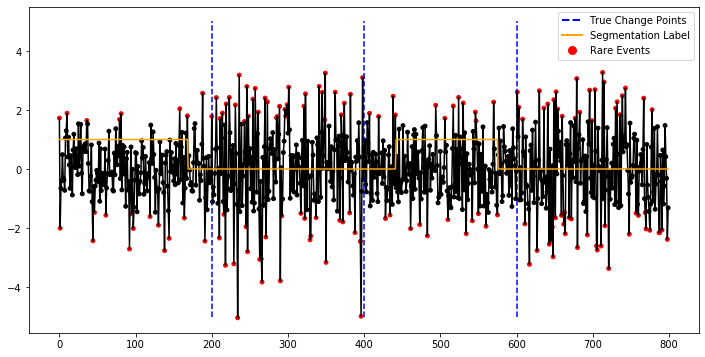}
\caption{Independently Normal distributed process with $\mu=0$ and $\sigma=1$ or $1.5$ varying over time. The vertical dashed line indicates the real change points; the yellow solid line indicates estimated segmentation label; red dots indicate the events of interest}
\label{fig.illustration}
\end{figure}

\begin{figure}[!h]
\centering
\includegraphics[width=5.2in]{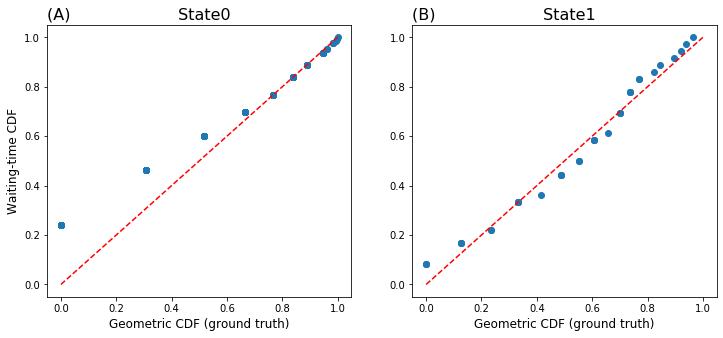}
\caption{P-P plot for the geometric distribution with true parameter versus empirical waiting time between successive events; (A) within ``state0''; (B) within ``state1''}
\label{fig.illustration_pp}
\end{figure}

A toy dataset is simulated to illustrate how the algorithm works. Independent Normal data points are simulated with mean 0 but time-varying variance. $\sigma=1$ when time $T \in [1,200] \bigcup [400,600]$, denoted as ``state1'', and $\sigma=1.5$ for the rest of time, denoted as ``state0''. The purpose here is to discover the underlying switching pattern of $\sigma$. The simulated time series is shown in Figure~\ref{fig.illustration}. A pair of thresholds $\alpha$ and $\beta$ is chosen as cutting lines to mark extremely large values (red dots). After that, the limiting distribution of waiting time between successive extreme events is analyzed, and segmentation is done via model selection with AIC. P-P plots in Figure~\ref{fig.illustration_pp} show a goodness-of-fit for the waiting time variables in both regions. And the segmentation result (yellow line) can almost perfectly capture the true dynamic pattern of $\sigma$.

There are still at least three shortcomings for this approach: (i) independence of returns is assumed without considering time dependence; (ii) decoding in multivariate settings yet to be developed; (iii) a more data-driven way to define an ``event'' is required. 

Here, we study this non-parametric region switching model under the assumption that ${X_t}$ are independent. Such independence assumption is practically needed for computational purposes, though it might be often violated in real settings. Indeed, this assumption allows us to connect the asymptotic conclusion with real data analysis. Results may also be useful when the assumption is slightly violated. We later discuss modifications to our proposed computational approach such that it could accommodate small degree of violation of this independence assumption. For the last two shortcomings, a novel decoding method is proposed to discover the stochastic dynamic among multivariate time series in Section3 and Section4.

\section{Multivariate Decoding}

\subsection{The Method}
Following the discussion of the excursion process, we shall extend the strategy from one dimension to multivariate. In single-dimensional time series, we consider volatility as a temporal aggregation of absolute large returns, so that extreme returns can be marked with appropriate choice of $\alpha$ and $\beta$ in \eqref{eq:cut}, and then
the dynamic pattern is revealed by decoding the resultant 0-1 sequence. However, without any clear definition of an event of interest, it raises a problem in multivariate settings. The event should be defined to reflect the dependence or at least local dependence of the multiple time series, for example, to mark data points contained in a pre-determined Euclidean subarea in $\mathbb{R}^p$. Intuitively, the subarea is of most interest if it contains data points exclusively from one underlying hidden state. A time period is currently under the control of this state if the subarea-specific events emerge chronologically in a high frequency.

Motivated by the idea of exploring local dependence, a new encoding and decoding approach is proposed as follows. In the encoding phase, a series of (rough) Euclidean ``ball'' is generated in $\mathbb{R}^p$ to mark points of interest, and so a series of 0-1 binary sequences can get returned. In the decoding phase, we treat the information of dynamics obtained from each ``ball'' as a feature and aggregate all pieces of information as one. The global pattern is ultimately discovered by clustering time points with similar feature sets.

Consider multivariate time series $\{X_t\}_{t=1}^{N}$. Let $B^{(v)}$ be the $v\textit{-th}$ ``ball'' with pre-fixed boundary. A new excursion process is defined by,

\begin{equation}\label{eq:ball}
E_t^{(v)} = \begin{cases} 1 & \quad X_t\in B^{(v)}\\
0 & \quad Otherwise \end{cases}
\end{equation}
Under the assumption of Theorem~\ref{th1}, the waiting time between two successive 1's in $E_t^{(v)}$ converges to a geometric distribution. The emission probability of 1's given hidden state $S_j$ now becomes 
\begin{equation}
p_{j}^{(v)}=\int_{B^{(v)}} dF_{j}
\end{equation}
where $F_{j}$ is the conditional CDF given $S_j$. A series of alternating hidden regions, for example $(S_1, S_1, S_2, S_1, ...)$, can be computed in model selection, and the corresponding region-based permission probability, which is $(p_{1}^{(v)}, p_{1}^{(v)}, p_{2}^{(v)}, p_{1}^{(v)}, ...)$ in this example, can be estimated by MLE of geometric distribution. Denote the N-length estimated probability vector as $\hat{p}^{(v)}$ which is the feature generated by $B^{(v)}$. Iteratively generate subarea $B^{(v)}$ for $v=1,2,...,V$, then $V$ resultant features can get obtained.

Note that $\hat{p}^{(v)}$ could be a vector with only a single value if the true permission probability are comparable given different hidden regions, for example, when $\int_{B^{(v)}} dF_{0} \cong \int_{B^{(v)}} dF_{1}$. In this case, features are less relevant or even redundant. In contrast, features may contain significant information about the dynamics when $\int_{B^{(v)}} dF_{0}$ differs a lot from $\int_{B^{(v)}} dF_{1}$.

\subsection{Ball Generation}

To make features more representative and less correlated, the ``balls'' should ideally be generated mutually disjointed and samples should get selected only once. In real data analysis with finite samples, it is neither efficient nor effective to determine the boundary for each ``ball'', especially when the dimension is high. Instead, we turn to select a fixed proportion of samples at each iteration and make each group of the samples less overlapping. 

To generate fewer overlapping sample groups, cluster analysis can be applied for the purpose. K-Means would be the most appropriate method due to its scalability and property of getting relatively balanced clusters. Assume $V$ clusters get returned via K-Means, then a rough ``ball'' can be generated by searching for $M$ nearest neighbors starting from the centroid of each cluster. The reason that we fix the size is to make sure there is enough data selected in each cluster. There is actually a tradeoff between the sample size of recurrent time and the magnitude of the excursion. We will keep using the proportion that is advocated in one-dimensional settings, say $\frac{M}{N}=0.1$. As a result, $V$ subarea gets returned, and each includes exactly $M$ data points. The $V$ is chosen very large in practice, say 100, so a sample is chosen 10 times on average. Here, we lose less information but via involving more correlated features. 

Let $\mathbb{X}=[X_1, X_2, ..., X_N]^T$ be a $N \times p$ matrix that records the time series $\{X_t\}_{t=1}^N$ where $X_t \in \mathbb{R}^p$. The feature generation algorithm is described in Algorithm~2. In the end, we simply stack all the features into a $N \times V$ matrix $\mathbb{P}=[\hat{p}^{(1)}, \hat{p}^{(2)}, ..., \hat{p}^{(V)}]$ as the output. The next task is resolved by feature selection or feature weighting techniques discussed in the next section.

\noindent\rule{12.5cm}{0.8pt}\\
\textbf{Algorithm~2} Feature Extraction\\
\rule{12.5cm}{0.4pt}\\
Input: Data matrix $\mathbb{X}$\\
1. Apply K-Means to $\mathbb{X}$, and get $V$ cluster centroids $C_1, C_2, ..., C_V$.\\
2. Loop: cycle through every $C_v$\\
\indent a. Search for its $M$ nearest neighbors in $\mathbb{X}$, denoted as $B^{(v)}$\\
\indent b. Generate a 0-1 excursion process via \eqref{eq:ball}, denoted as $\{E_t^{(v)}\}_{t=1}^N$\\
\indent c. Apply Algorithm~1 to $\{E_t^{(v)}\}_{t=1}^N$ to get emission probability $\hat{p}^{(v)}$\\
3. Stack all $\hat{p}^{(v)}$'s into a $(N \times V)$ matrix $\mathbb{P}=[\hat{p}^{(1)}, \hat{p}^{(2)}, ..., \hat{p}^{(V)}]$,\\
\indent and record $B^{(v)}$'s in a set $\mathbb{B}=\{B^{(1)},B^{(2)},...,B^{(V)}\}$.\\
Output: confusion matrix $\mathbb{P}$ and set $\mathbb{B}$\\
\rule{12.5cm}{0.8pt}\\

\section{Feature Weighting}

The decoding result is finally achieved by clustering similar time points in $\mathbb{P}$. We will use K-Means as an example to illustrate the idea. K-Means minimizes the sum of within-cluster error via iteratively assigning each object by its closest centroid and updating each centroid consequently.
Define $Y_{iv}$ is the $v$-th feature in the $i$-th sample, for $v=1,2,...,V$, and $C_j$ is the centroid of the $j$-th cluster $S_j$, for $j=1,2,...,k$. The optimization problem can be specified as to minimizing the following quantity,

\begin{equation}\label{eq:opt}
\sum_{j=1}^k \sum_{i\in S_j} \sum_{v=1}^V D(Y_{iv}, C_{jv})
\end{equation}
where $D(.)$ is a metric.

As what is discussed before, features may have different degrees of relevance, but K-Means treats every single feature equally, regardless of the actual relevance. As a consequence, clustering results could be greatly biased by the irrelevant features, while the more relevant features are overwhelmed. 
This weakness can be resolved by feature selection or feature weighting which is discussed as follows. 

The research in feature weighting of clustering can be traced back to 1984. Different from feature selection, feature weighting approaches usually lead to better performance by iteratively conducting clustering and adjust feature weights based on the result in the last step. A survey on feature weighting of K-Means is available in \cite{19}. Typically, the goal is to minimize the within clustering dispersion by updating the feature weight $w_v$ for feature position $v$. The optimization problem \eqref{eq:opt} is then rewritten as,

\begin{equation}
\sum_{j=1}^k \sum_{i\in S_j} \sum_{v=1}^V w_v D(Y_{iv}, C_{jv})
\end{equation}

Usually, $w_v$ is set so that $\sum_{v=1}^V{w_v}=1$. Note that $w_v$ may also vary in different clusters, and metric $D(.)$ can be generalized to non-Euclidean distance, like Minkowski's \cite{23}, but they are beyond our focus in this paper.

\subsection{Related Works}

Feature Weight Self-Adjustment mechanism(FWSA) \cite{20} is designed to adjust feature weight to simultaneously minimize the separations within clusters and maximize the separations between clusters. The importance of a feature to the clustering quality is measured based on a function of sum of separations within clusters, denoted as $a_v$, and sum of separations between clusters, denoted as $b_v$, and feature weight is updated, iteratively,

\begin{equation}\label{eq:w1}
w_v^{(c+1)} = w_v^{(c)} + \eta\, ( w_v^{(c)} - \frac{b_v^{(c)}/a_v^{(c)}}{\sum_u b_u^{(c)}/a_u^{(c)}} )
\end{equation}
where $c$ indicates the current step, and $c+1$ is the next step; $\eta$ is the learning rate. The updated weight vector still sums up to 1. In the original paper, $\eta$ is set as 0.5. FWSA mechanism significantly improves clustering quality in experiments. In addition, it takes considerable advantage that no extra parameter is required to be specified.


The second method weights features according to mutual information. As Shannon Entropy is widely used as criteria of clustering quality, its variant, Mutual information, measures the amount of information obtained about the clusters which can be interpreted through another random variable. The first method quantifies the degree of relevance for a single feature by the normalized mutual information between clusters and features. 

The minimum of MI is 0 if a particular feature does not contribute any new information about what its cluster might be. Maximum mutual information is reached when a feature can perfectly recreate the clusters. A drawback of MI is that a feature with numerical value has to be categorized before applying the discrete-version formula, and entropy tends to increase with the number of categories. The Normalized Mutual Information(NMI) solves the problem by standardizing the MI number always between 0 and 1. Fortunately, no extra binning is required in matrix $\mathbb{P}$ since each feature is a sequence of discrete probability numbers. Feature weights are updated based on the idea that more relevant features to the current clustering result weights more than redundant features. 

\begin{equation}\label{eq:w2}
w_v^{(c+1)} = w_v^{(c)} + \eta\, ( w_v^{(c)}-\frac{NMI(L^{(c)}, Y_{.v}^{(c)})}{\sum_u NMI(L^{(c)}, Y_{.u}^{(c)})} )
\end{equation}
where $Y_{.v}=(Y_{1v},Y_{2v},...,Y_{Nv})^T$, $L^{(c)}$ is the cluster labels returned at the current step, and $\eta$ is the learning rate. As an unsupervised algorithm, the quality of clustering may get out of control, especially when signal-to-noise ratio is relatively low, so the noise may get exaggerated in the iteration.

\subsection{Feature Weighting Clustering for Decoding}

A new feature-weighting clustering algorithm is designed for the decoding procedure. As is claimed in one-dimensional settings, the decoding result is reliable if the true permission rates difference between two hidden states is large. It is the reason that $\alpha$ and $\beta$ in \eqref{eq:cut} are tuned to enlarge the difference between the tailedness of underlying distributions. Inspired by this idea, the feature importance can also be measured by the estimated permission rate delta. 

Recall the time series data $\{X_t\}_{t=1}^N$. In a iterative fashion, let $L^{(c)}_t$ be the cluster label for data point $X_t$ in the current step. When $k=2$, $L^{(c)}_t$ only takes two values corresponding to two hidden states, say ``state0'' and ``state1''. Denote the $v$-th feature is generated by a selection area $B^{(v)}$, then permission probability $p_j^{(v)(c)}$ upon $B^{(v)}$ given hidden state $j$ can be further estimated by,

\begin{equation}
\hat{p_j}^{(v)(c)}=\frac{\sum_{t=1}^N 1\{L_t=j, X_t \in B^{(v)}\}}{\sum_{t=1}^N 1\{L_t=j\}}
\end{equation}
Especially when $k=2$, the feature importance for feature $v$ can be quantified based on $|\hat{p_1}^{(v)(c)}-\hat{p_0}^{(v)(c)}|$. The greater the absolute difference, the more important feature $v$ is. The feature weight can be simply updated by,

\begin{equation}\label{eq:wa}
w_v^{(c+1)} = w_v^{(c)} + \eta\, ( w_v^{(c)} - \frac{|\hat{p_1}^{(v)(c)}-\hat{p_0}^{(v)(c)}|}{\sum_u |\hat{p_1}^{(u)(c)}-\hat{p_0}^{(u)(c)}|} )
\end{equation}

Without any prior information, let's assume the size of the two hidden states is balanced. Then, the estimated delta is simply measured by the proportion of the two cluster labels in $B^{(v)}$. The more purity of cluster in $B^{(v)}$, the more important feature $v$ should be. It actually enlighten us to look at the Shannon entropy in $B^{(v)}$ as a smooth approximation to $|\hat{p_1}^{(v)}-\hat{p_0}^{(v)}|$. 

\begin{equation}\label{eq:wb}
w_v^{(c+1)} = w_v^{(c)} + \eta\, ( w_v^{(c)} - \frac{e^{-H(B^{(v)})^{(c)}}}{\sum_u e^{-H(B^{(u)})^{(c)}}} )
\end{equation}
where $H(B^{(v)})^{(c)}$ denote the Shannon entropy of cluster labels of data points in $\{B^{(v)}\}$ at the current step. The feature weight is measured by one minus the purity of clusters in $B^{(v)}$. Note that the entropy-type feature weighting procedure can be easily generalized when $k>2$. Moreover, it takes advantages in geometric interpretation, which is illustrated in the simulation study.

\section{Simulation Experiments}

\subsection{Independent Processes}

Independent Bivariate Normal processes are simulated with mean 0 and 2 types of covariance matrix varying over time. The data is generated with a covariance matrix $Cov_0$ in a short period of time, then switching to the other matrix $Cov_1$ for a period and switching back, so on and so forth. Each short period indicates a state hidden behind the time series, and the conditional distribution given a state is identical. There are 10 alternating periods in total, and the time length for each period is uniformly distributed by $Unif([200,400])$.

\begin{figure}[!h]
\centering
\includegraphics[width=5.2in]{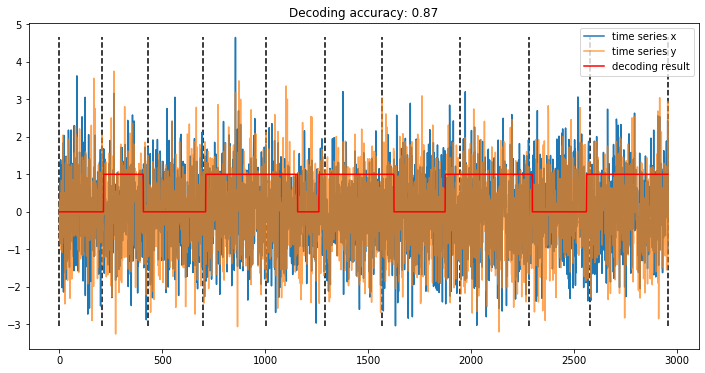}
\caption{Dataset simulated from bivariate Gaussian ``Case1''; vertical dashed line indicates the true change points; red solid line reflects the segmentation result via \eqref{eq:wb}}
\label{fig.data1}
\end{figure}

Consider 5 different simulation scenarios, named ``Case1'', up to ``Case5''. The detail about the simulated covariance matrix is reported in \hyperref[appn]{Appendix}. A confusion matrix is firstly obtained via feature generation (Algorithm~2), and then feature weighting K-Means is applied to clustering time points in hidden states. Figure~\ref{fig.data1} illustrates a decoding result in ``Case1''. It shows that the underlying dynamics pattern can be almost perfectly discovered although some stamps around lag 1000 are misclassified (accuracy is 0.87).

Four feature weighting clustering algorithms described in \eqref{eq:w1}, \eqref{eq:w2}, \eqref{eq:wa}, and \eqref{eq:wb} are compared. For the convenience of comparison, clustering accuracy is calculated and used to measure the quality of decoding. Denote the first feature-weighting algorithm in \eqref{eq:wa} as ``MethodA'', and the second one in \eqref{eq:wb} as ``MethodB''. Dataset is simulated for at least 100 times, and the decoding accuracy is reported in Table~\ref{table1}.

\begin{table}[h]
\centering
\caption{Decoding Accuracy}
\label{table1}
\begin{tabular}{@{}lrrrc@{}}
\hline
Simulation\\ 
Case      & \multicolumn{1}{c}{FWSA}            & \multicolumn{1}{c}{NMI}             & \multicolumn{1}{c}{MethodA}                  & \multicolumn{1}{c}{MethodB}                  \\ \hline
Case1 & 0.8048 (0.1046)         & 0.8021 (\textbf{0.0960})         & 0.8145 (0.1111)          & \textbf{0.8286} (0.1026)         \\ 
Case2 & 0.9377 (0.0190) & 0.9415 (0.0186) & \textbf{0.9502} (0.0145) & 0.9489 (\textbf{0.0145})          \\ 
Case3 & 0.9354 (0.0192) & 0.9389 (0.0170) & 0.9480 (0.0176)          & \textbf{0.9493} (\textbf{0.0146}) \\ 
Case4 & 0.9213 (0.0218) & 0.9249 (0.0217) & 0.9378 (\textbf{0.0142})          & \textbf{0.9390} (0.0142) \\ 
Case5 & 0.8920 (0.0490) & 0.8764 (0.0541) & 0.9030 (\textbf{0.0476})          & \textbf{0.9110} (0.0579) \\ \hline
\end{tabular}
\end{table}

\begin{figure}[!h]
\centering
\includegraphics[width=5.0in]{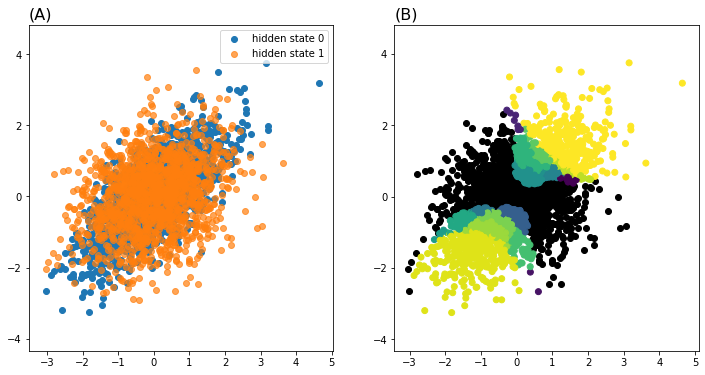}
\caption{Dataset simulated from bivariate Gaussian ``Case1''; (A) scartterplot from two hidden states; (B) data points are plotted in back; ``balls'' with high weights are painted in different color}
\label{data1_scatter}
\end{figure}

\begin{figure}[!h]
\centering
\includegraphics[width=5.0in]{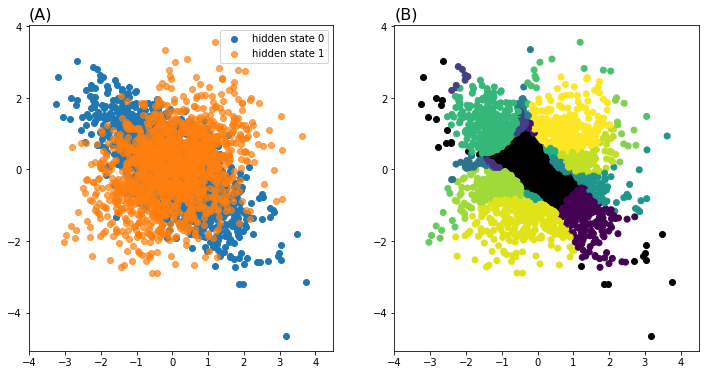}
\caption{Dataset simulated from bivariate Gaussian ``Case2''; (A) scartterplot from two hidden states; (B) data points are plotted in back; ``balls'' with high weights are painted in different color}
\label{data2_scatter}
\end{figure}

It turns out that the feature weighting methods are adapt to the decoding framework well. ``MethodB'' that weights features according to the entropy of each Euclidean ``ball'' outperforms others. In ``Case1'', the join distribution given a hidden state is Gaussian with a unit variance but different correlations. The join distribution in the two states can be visualized from Figure~\ref{data1_scatter}(A). ``Balls'' with relatively high feature weights are highlighted in Figure~\ref{data1_scatter}(B). It looks that the algorithm is trying to pay more attention to the ``balls'' located in the right-up and left-bottom corners, in which the distributions differ a lot, see Figure~\ref{data2_scatter}. While in ``Case2'', weights are concentrated to ``balls'' located around the four corners. It claims that our new feature weighting strategy can truly find out the key difference between the joint distributions, and ``balls'' with high weights play a significant role in detecting the distribution changes.

\subsection{Serial Dependent Processes}

In this section, we discuss some extensions to our approach when serial dependence is present in the time sequence. This problem is related to change point detection in time series models. To detect structural breaks in variance, authors in \cite{21} studied cumulative sums of squares in case of independent sequence. Later on, the test statistic is modified by looking into the stability breaks of the autocovariance function $\gamma(r)=E[X_{t}X_{t+r}]$ where $r$ is the time lag \cite{22}.

Motivated by the idea, we extend the multivariate decoding procedure to a single time series with weak serial dependence. A multivariate process is made up by coupling time point with its $r$-lags, say $\{Z_t\}_t=\{(X_t, X_{t+1},... X_{t+r})\}_{t=1}^{N-r}$. We suppose the $(r+1)$-dimensional variables can represent the covariance structure, and modify the decoding algorithm as follows. For the validation of the independence assumption, it is necessary to break the local dependence. Time sequence $\{Z_t\}_t$ is partitioned by $l$-length window, so $\floor{\frac{N-r}{l}}$ time pierces are obtained. Time points in each window are then randomly permuted and denote the new sequence as $\{\Tilde{Z}_t^l\}_t$. The choice of $l$ could be very tricky. A too small $l$ has nothing to do with breaking the dependence, while a too large $l$ tremendously destroys the true dynamic pattern. We find $l$ around 30 is proper given that the size of a hidden region is at least 300.

Datasets are simulated based on AR(1) and AR(2) models. Independent standard normal variables were used as innovations. In AR(1) settings, parameters are set $\phi|state0 = 0.3$ and $\phi|state1 = 0.7$ given hidden state ``state0'' and ``state1'', respectively. In AR(2), the pair of parameters is $(\phi_1, \phi_2)|state0 = (0.3, 0.2)$ and $(\phi_1, \phi_2)|state1 = (0.5, 0.3)$. The switching pattern of the hidden states is generated in a fashion similar to that in Section 5.1. We choose $r=1$ and $2$ to make up new time sequence $\{Z_t\}_t$ in AR(1) and AR(2) settings, respectively. The average decoding accuracy for simulation in AR(1) is 0.7791 (0.0861), and 0.8045 (0.0705) in AR(2).

\begin{figure}[!h]
\centering
\includegraphics[width=5.2in]{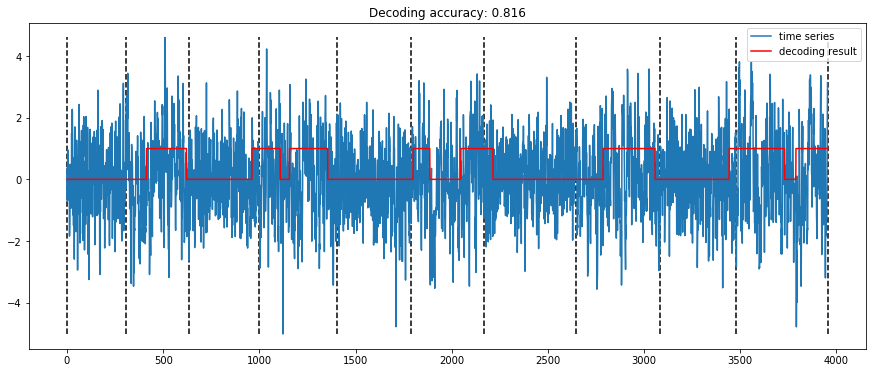}
\caption{Dataset simulated from AR(1) with $\phi|state0 = 0.3$ and $\phi|state1 = 0.7$}
\label{AR1}
\end{figure}

\begin{figure}[!h]
\centering
\includegraphics[width=5.2in]{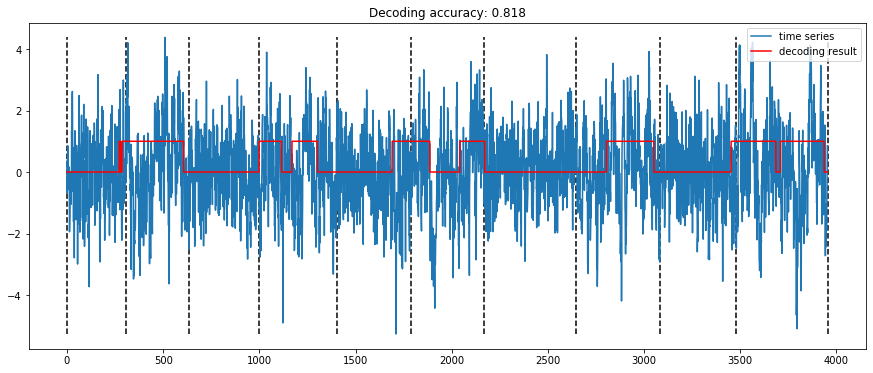}
\caption{Dataset simulated from AR(2) with $(\phi_1, \phi_2)|state0 = (0.3, 0.2)$ and $(\phi_1, \phi_2)|state1 = (0.5, 0.3)$}
\label{AR2}
\end{figure}

\section{Real Data Application}

\subsection{Triplet Time Series}

The relationship between returns, trading volume, and transaction numbers has been received great amounts of attention in finance. Under one old Wall Street adage that ``it takes volume to move prices'', volume had been increasingly used as a cause of return volatility. It can be explained that volume can reflect the extent of disagreement about a security's value in stock price. However, it would be modified later that it is the number of trades but their sizes that generate volatility \cite{24}. It would also be shown that to recover normality in asset returns, the number of trades is a better time change than the traditionally used trading volume.

It is claimed in \cite{16} that direct modeling may have difficulty capturing the intricate dynamic structure, especially given the lack of goodness-of-fit in dynamic linear regression. A nonparametric approach was advocated to explore each of the three dimensions separately by segmenting volatility and non-volatility states, and then combine them to reflect a single stock dynamics. However, the idea of divide-and-conquer may fail to capture the real association among the three but be biased by the integrated microstructure noises.

In the experiment, we track the 3-dimensional time series of a single stock from S$\&$P500. The log return, volume, and transaction number at every 1-min interval are recorded. To mitigate the influence of activities near opening and closing, We truncate the transaction time from 10am to 4pm, so there are 360 data points per business day. Again, no prior knowledge about the stochastic mechanism needs to be assumed. Via our proposed method, the time axis is segmented into equilibrium and off-equilibrium periods to represent the latent state-space trajectory underlying the single stock's dynamics.

\begin{figure}[!h]
\centering
\includegraphics[width=5.2in]{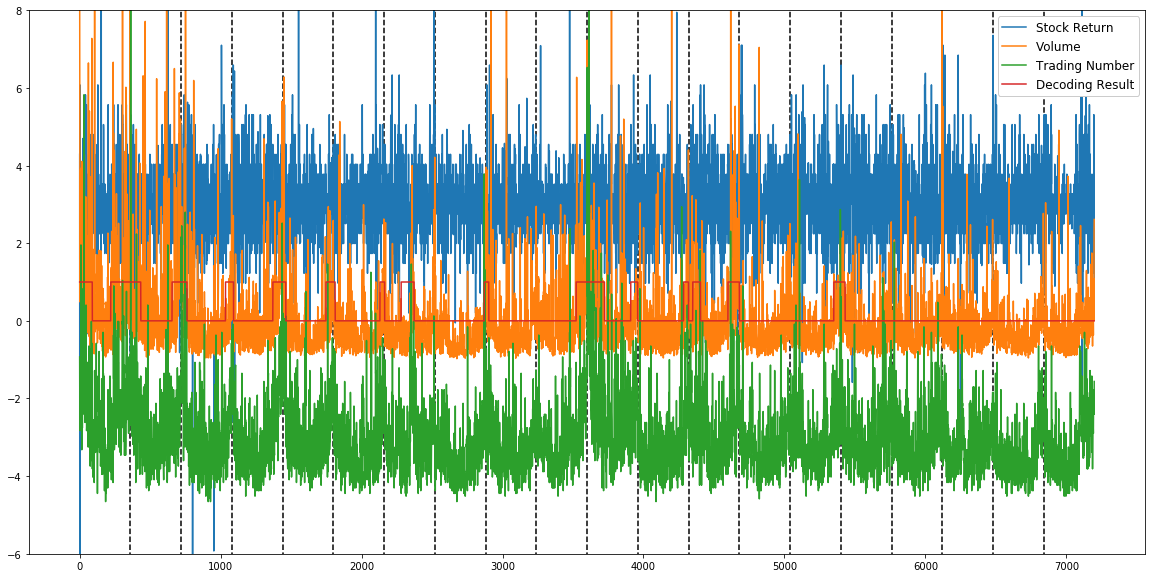}
\caption{Trivariate time series of IBM}
\label{fig.IBM}
\end{figure}

Figure~\ref{fig.IBM} shows minutely trivariate time series of IBM in January 2006. Each of the dimensions is standardized to have a mean 0 and standard deviation of 1. A constant is added or subtracted to returns and trading numbers for better visualization so that the 3 time series are clearly viewed in one panel. The vertical dashed line indicates a date change. It shows that volume and trading number are highly correlated. They would rhythmically go up and down simultaneously. The decoding result (0-1 sequence) obtained by our method is plotted in a red line, which represents the two hidden states switching throughout the whole period. It looks that the segmentation can successfully capture the time when volume and trading number both increase heavily, see state code ``1''. If the increment is not that much, it is marked as in equilibrium state, see the right part in Figure~\ref{fig.IBM}.

\begin{figure}[!h]
\centering
\includegraphics[width=5.2in]{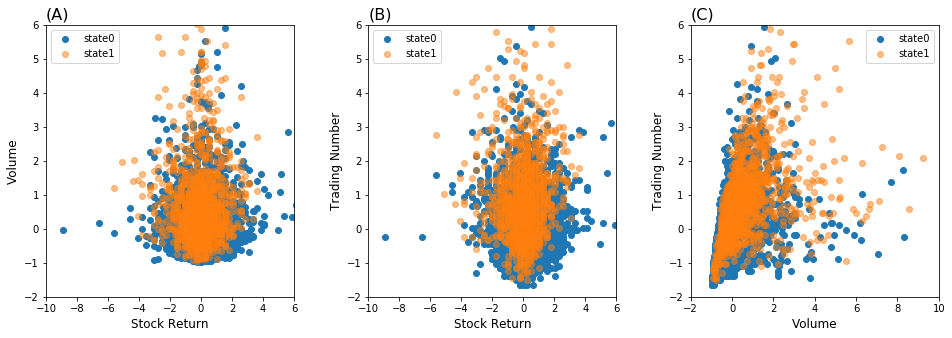}
\caption{2-D scatterplot for IBM: (A) returns v.s volume; (B) returns v.s trading numbers (C) volume v.s trading numbers}
\label{fig.IBM_scatterplot}
\end{figure}

\begin{figure}[!h]
\centering
\includegraphics[width=5.2in]{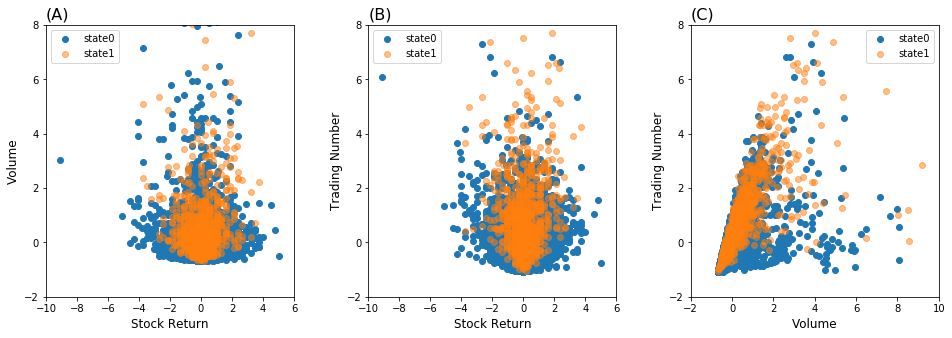}
\caption{2-D scatterplot for ADBE: (A) returns v.s volume; (B) returns v.s trading numbers (C) volume v.s trading numbers}
\label{fig.ADBE_scatterplot}
\end{figure}

The next question is, what is the association among these 3 time series given different hidden states? 2-D scatterplots in Figure~\ref{fig.IBM_scatterplot} can roughly illustrate the answer. The correlation between volume and trading number is much higher in ``state1''. The surprising pattern is that the corresponding stock returns in the same period have a much lower deviation than that in ``state0''. That is to say, stock return tends to stay non-volatile once volume and trading number are significantly going up together. This phenomenon is shown more clearly in Adobe's stock, see Figure~\ref{fig.ADBE_scatterplot}. Our findings contradict the previous argument that volatility is highly correlated to volume or trading numbers.

\subsection{Multivariate Returns}

\begin{figure}[!h]
\centering
\includegraphics[width=5.2in]{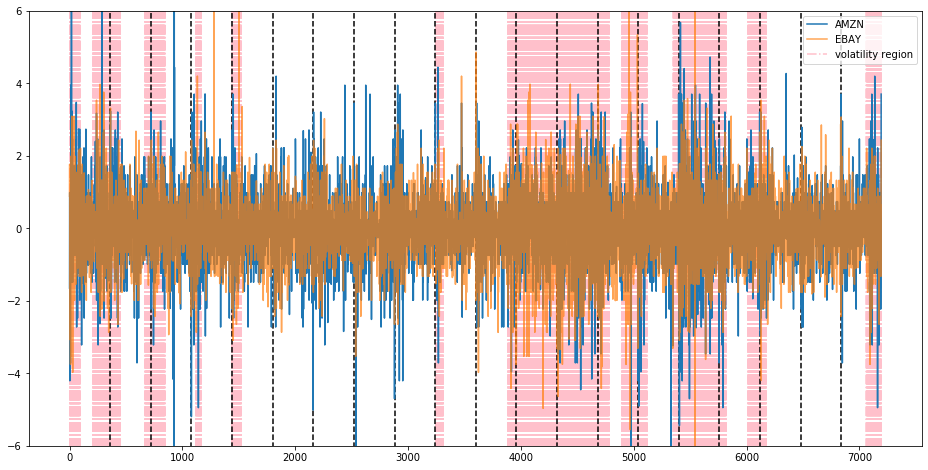}
\caption{Bivariate returns of Amazon and Ebay in January 2006 }
\label{fig.internet_retail}
\end{figure}

\begin{figure}[!h]
\centering
\includegraphics[width=5.2in]{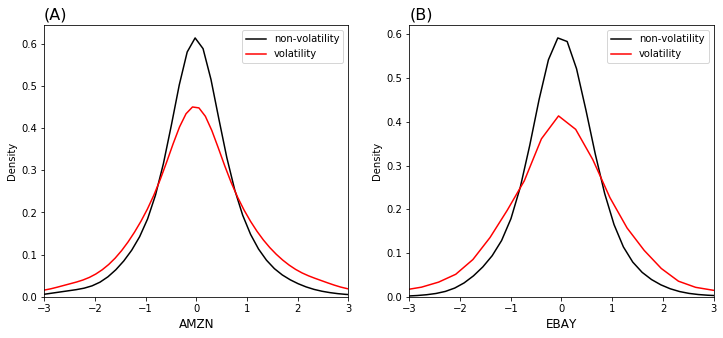}
\caption{Kernel density estimation for data points on volatility and non-volatility region; (A) Amazon; (B) Ebay}
\label{fig.internet_retail_pdf}
\end{figure}

In this experiment, we apply the decoding approach to discover the time-varying dependence among bivariate and multivariate stock returns. In the first example, a pair of indexes is chosen from one of the categories of S$\&$P500 based on Global Industrial Classification Standard(GICS). For example, `Amazon' and `Ebay' are coupled together as a pair of representatives for internet retails. The price returns are calculated in 1-min time interval. The volatility segmentation result is shown in Figure~\ref{fig.internet_retail}. Volitility state shows up rhythmically every day in the first business week, and then the frequency tends to disappear in the second week. The return fluctuates even severely in the third week and then goes back to the first state in the end. Kernel density estimations for returns in volatility and non-volatility stages are plotted separately in Figure~\ref{fig.internet_retail_pdf}. It shows that both distributions have their tails heavier when in volatility stage.

\begin{figure}[!h]
\centering
\includegraphics[width=5.2in]{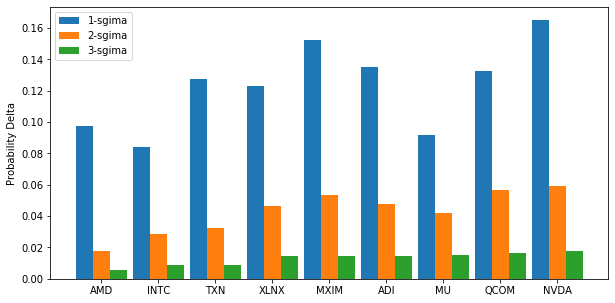}
\caption{Heavy-tailedness delta between volatility and non-volatility; 9 indexes from left to right is `AMD', `INTC', `TXN', `XLNX', `MXIM', `ADI', `MU', `QCOM', and `NVDA'}
\label{fig.prob_delta}
\end{figure}

In the second example, we pick up 9 semiconductor indexes from S$\&P$500 and segment the time axis into volatility and non-volatility regions. To measure the heavy-tailedness, we calculate the probability with which return $X$ goes beyond the $z$-standard deviation limits, for $z=1,2,3$,
$$P(X<z\,\sigma) + P(X>z\,\sigma)$$ 
The heavy-tailedness is calculated for volatility and non-volatility, respectively. The delta values between volatility and non-volatility is reported in Figure~\ref{fig.prob_delta}. All the positive delta values indicate that the 9 indexes would have a heavier tail simultaneously when in the volatility period, but the heavy-tailedness is quite different. For example, `Advanced Micro Devices(AMD)' and `Intel(INTC)' have relatively stable returns when in volatility; while returns of `Qualcomm(QCOM)' and `Nvida(NVDA)' fluctuate more heavily.

\section*{Conclusion}

In the paper, we try to break down the complicated model framework and to directly investigate the volatility dynamic patterns underlying multivariate stock time series. A feature engineering strategy is proposed from feature extraction to feature weighting, and our clustering results can successfully detect the switching region in which the nonlinear dependence differs a lot. In the real data experiment, we revised the former claim on the relationship among returns, trading volume, and transaction numbers, and measure the association in multiple return time series. Despite the weakness in modeling long-term serial dependence and forecasting, the data-driven approach established a platform to study distributional heterogeneity, which is commonly observed in reality. In the future, it can incorporate time series models, like GARCH, to perform a more detailed analysis.

\newpage

\begin{appendix}
\section*{Appendix}\label{appn}

Denote the two hidden states as ``state0'' and ``state1'', and their corresponding covariance matrix ``$Cov_0$'' and ``$Cov_1$'', respectively. In Section~5, datasets are simulated in 5 different cases described as following. 

\noindent\textbf{A. Case1}

\[ Cov_0= \begin{bmatrix}
1 & 0.3 \\
0.3 & 1 
\end{bmatrix} \]

\[ Cov_1= \begin{bmatrix}
1 & 0.7 \\
0.7 & 1 
\end{bmatrix} \]

\noindent\textbf{B. Case2}

\[ Cov_0= \begin{bmatrix}
1 & 0.3 \\
0.3 & 1 
\end{bmatrix} \]

\[ Cov_1= \begin{bmatrix}
1 & -0.7 \\
-0.7 & 1 
\end{bmatrix} \]

\noindent\textbf{C. Case3}

\[ Cov_0= \begin{bmatrix}
\sigma_1^2 & r*\sigma_1*\sigma_2 \\
r*\sigma_1*\sigma_2 & \sigma_2^2
\end{bmatrix} \]

\[ Cov_1= \begin{bmatrix}
\sigma_2^2 & r*\sigma_1*\sigma_2 \\
r*\sigma_1*\sigma_2 & \sigma_1^2
\end{bmatrix} \]

where $\sigma_1$=1, $\sigma_2$=1.5, r=0.6.

\noindent\textbf{D. Case4}

\[ Cov_0= \begin{bmatrix}
\sigma_1^2 & r*\sigma_1*\sigma_2 \\
r*\sigma_1*\sigma_2 & \sigma_2^2
\end{bmatrix} \]

\[ Cov_1= \begin{bmatrix}
\sigma_2^2 & r*\sigma_1*\sigma_2 \\
r*\sigma_1*\sigma_2 & \sigma_1^2
\end{bmatrix} \]

where $\sigma_1$=1, $\sigma_2$=1.5, r=0.2.

\noindent\textbf{E. Case5}

\[ Cov_0= \begin{bmatrix}
1 & 0.3 \\
0.3 & 1 
\end{bmatrix} \]

\[ Cov_1= \begin{bmatrix}
1 & -0.3 \\
-0.3 & 1 
\end{bmatrix} \]

\begin{figure}[!h]
\centering
\includegraphics[width=5.0in]{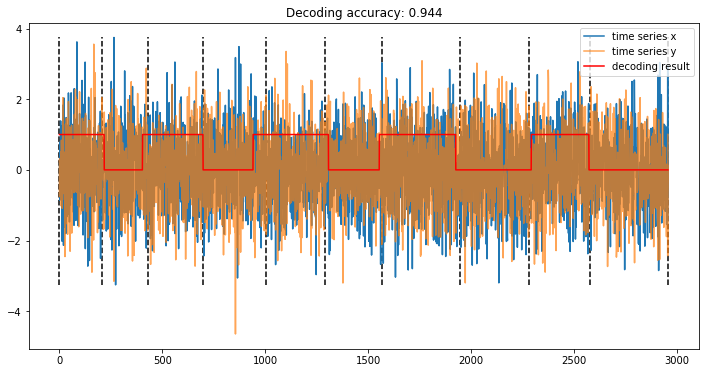}
\caption{Dataset simulated from bivariate Gaussian ``Case2''; vertical dashed line indicates the true change points; red solid line reflects the segmentation result}
\label{data2}
\end{figure}

\begin{figure}[!h]
\centering
\includegraphics[width=5.0in]{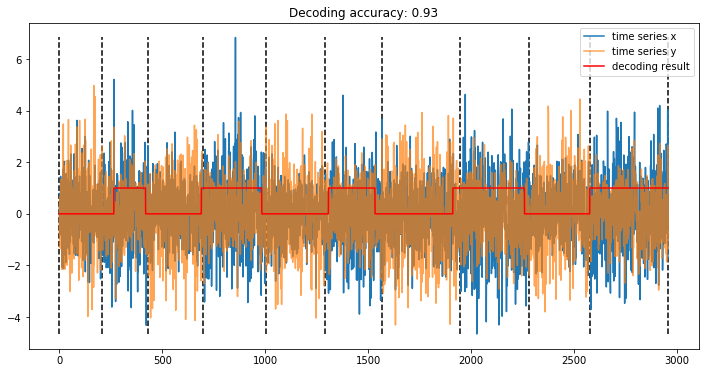}
\caption{Dataset simulated from bivariate Gaussian ``Case3''; vertical dashed line indicates the true change points; red solid line reflects the segmentation result}
\label{data3}
\end{figure}

\begin{figure}[!h]
\centering
\includegraphics[width=5.0in]{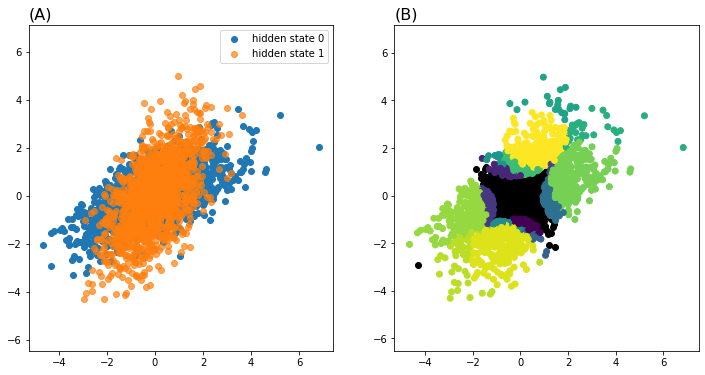}
\caption{Dataset simulated from bivariate Gaussian ``Case3''; (A) scartterplot from two hidden states; (B) data points are plotted in back; ``balls'' with high weights are painted in different color}
\label{data3_scatter}
\end{figure}

\begin{figure}[!h]
\centering
\includegraphics[width=5.0in]{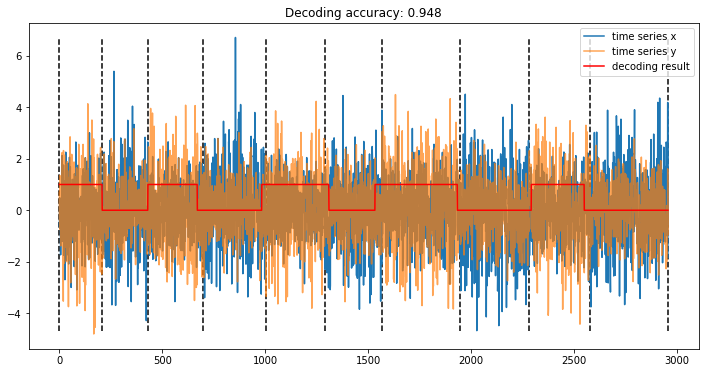}
\caption{Dataset simulated from bivariate Gaussian ``Case4''; vertical dashed line indicates the true change points; red solid line reflects the segmentation result}
\label{data4}
\end{figure}

\begin{figure}[!h]
\centering
\includegraphics[width=5.0in]{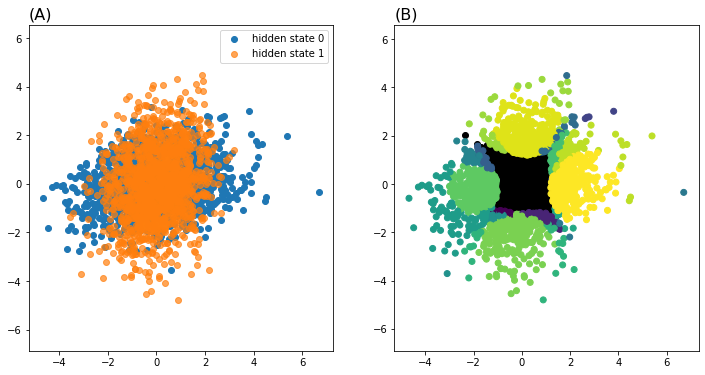}
\caption{Dataset simulated from bivariate Gaussian ``Case4''; (A) scartterplot from two hidden states; (B) data points are plotted in back; ``balls'' with high weights are painted in different color}
\label{data4_scatter}
\end{figure}

\begin{figure}[!h]
\centering
\includegraphics[width=5.0in]{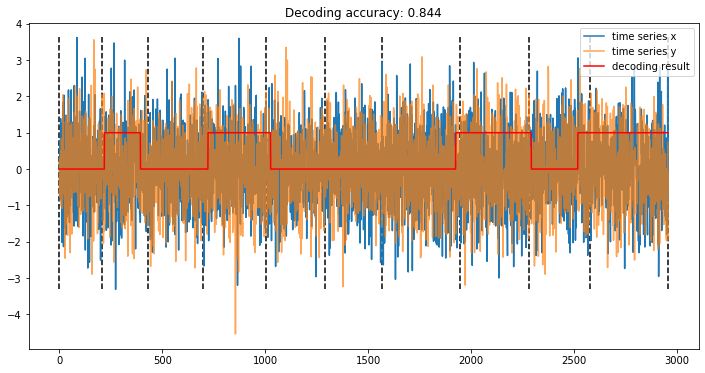}
\caption{Dataset simulated from bivariate Gaussian ``Case5''; vertical dashed line indicates the true change points; red solid line reflects the segmentation result}
\label{data5}
\end{figure}

\begin{figure}[!h]
\centering
\includegraphics[width=5.0in]{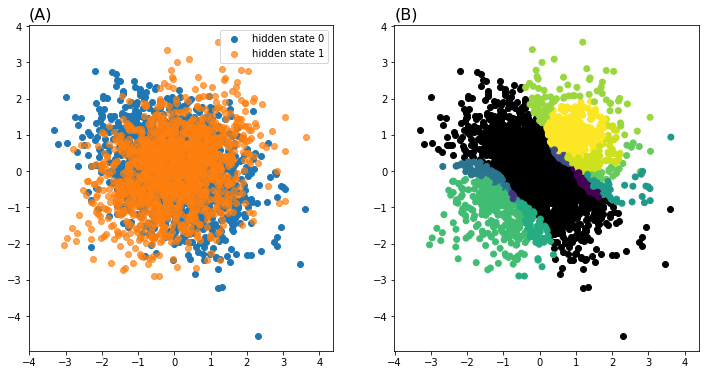}
\caption{Dataset simulated from bivariate Gaussian ``Case5''; (A) scartterplot from two hidden states; (B) data points are plotted in back; ``balls'' with high weights are painted in different color}
\label{data5_scatter}
\end{figure}

\begin{figure}[!h]
\centering
\includegraphics[width=5.2in]{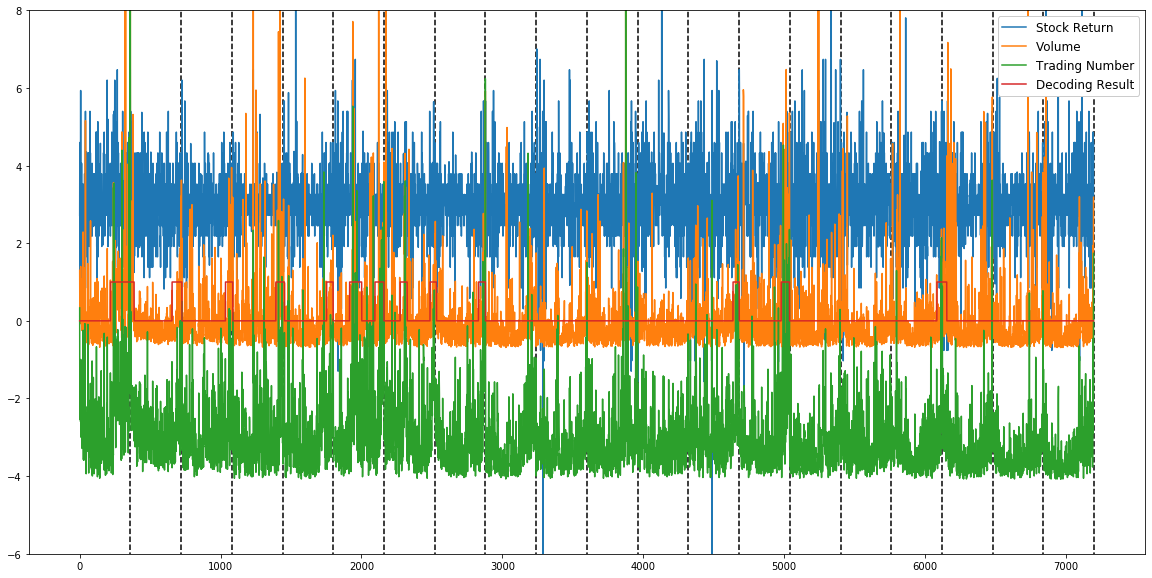}
\caption{Trivariate time series of ADBE}
\label{fig.ADBE}
\end{figure}

\begin{figure}[!h]
\centering
\includegraphics[width=5.2in]{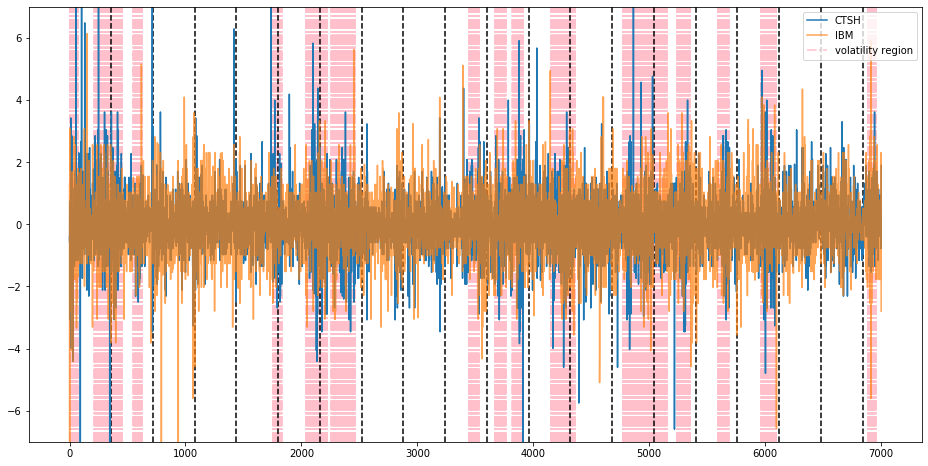}
\caption{Bivariate returns of CTSH and IBM}
\label{fig.it_consulting}
\end{figure}

\begin{figure}[!h]
\centering
\includegraphics[width=5.2in]{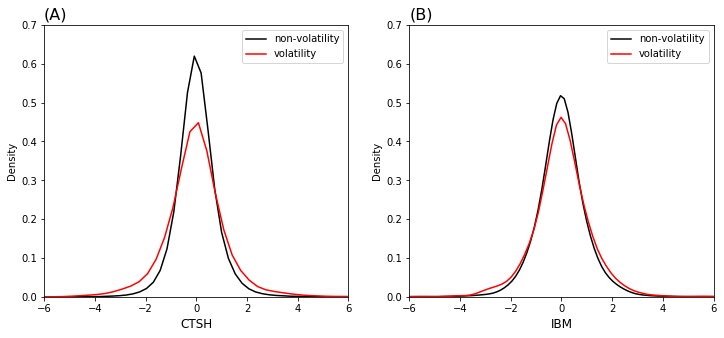}
\caption{Kernel density estimation for data points on volatility and non-volatility region; (A) CTSH; (B) IBM}
\label{fig.it_consulting_density}
\end{figure}

\begin{figure}[!h]
\centering
\includegraphics[width=5.2in]{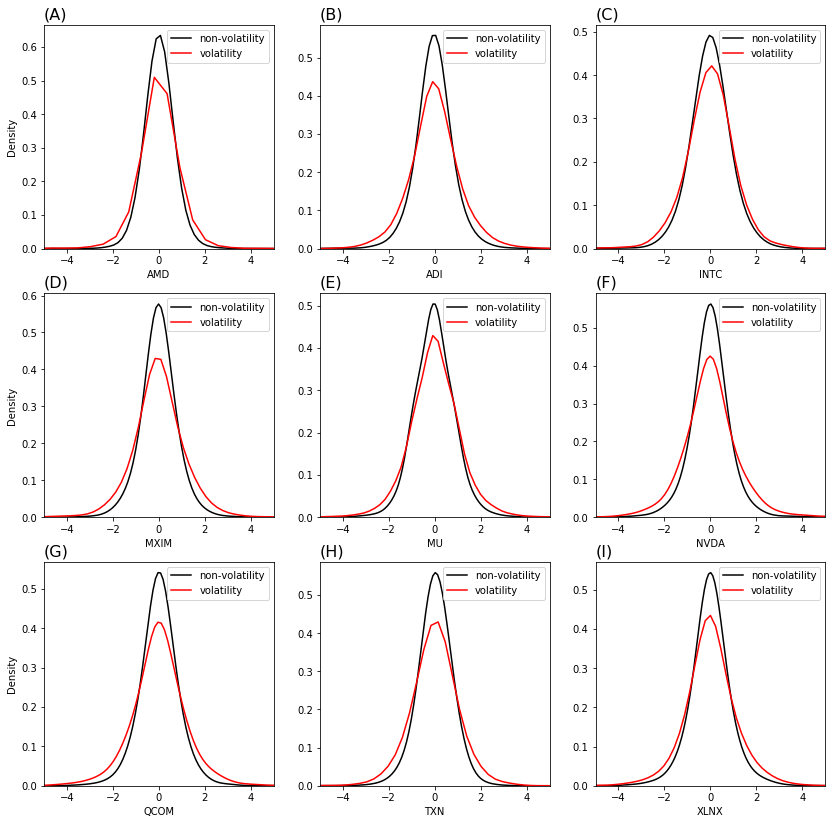}
\caption{Kernel density estimation for data points on volatility and non-volatility region for 9 semiconductor indexes}
\label{fig.semiconductor_density}
\end{figure}

\end{appendix}

\end{document}